# The role of current loop in harmonic generation from magnetic metamaterials in two polarizations

**Iman Sajedian[1,2], Inki Kim[2], Abdolnasser Zakery[1] and Junsuk Rho[2,3*]**

[1]*Department of Physics, College of Science, Shiraz University, Shiraz 71454, Iran*
[2]*Department of Mechanical Engineering, Pohang University of Science and Technology (POSTECH), Pohang 37673, South Korea*
[3]*Department of Chemical Engineering, Pohang University of Science and Technology (POSTECH), Pohang 37673, South Korea*
[*] *jsrho@postech.ac.kr*

In this paper, we investigate the role of current loop in the generation of second and third harmonic signals from magnetic metamaterials. We will show that the fact that the current loop in the magnetic resonance acts as a source for nonlinear effects and it consists of two orthogonal parts, leads to the generation of two harmonic signals in two orthogonal polarizations. Each of these parts is responsible for generating a harmonic signal parallel to itself. The type of harmonic signal is determined by the metamaterial's inversion symmetry in that direction. We believe that our claim, which was also observed in the experimental work of another group, will help improve the measurements of nonlinear effects from magnetic metamaterials.

**Introduction**

Considering the recent rapid growth of photonic science and the birth of newly designed photonic devices, the need for new photonic elements has been increasing. Ones of such elements are nonlinear metamaterials, which are artificially structured materials capable of generating nonlinear effects in very small thickness. Researchers have been trying to find metamaterials for generating nonlinear effects for the past decades, and some designs have been introduced for second harmonic generation[1-3]. Some others have used the nonlinear photonic elements such as diodes for generating nonlinear effects[4,5]. Other nonlinear effects like solitons are also seen in metamaterials[6], and tunable nonlinear metamaterials are also made by using liquid crystals[7].

In this paper, we show how magnetic metamaterials generate harmonic signals in two polarizations. Magnetic metamaterials produce their magnetic response by inducing a magnetic dipole in the opposite direction of the incident magnetic field. This magnetic dipole is created by a loop of current in the magnetic resonance of the metamaterial[8]. In this paper, we investigate the effect of this current loop in harmonic generation. This claim is supported by numerical analysis of nanostrips as a sample of magnetic metamaterials.

## Results

**Theory**

The optical properties of metamaterials can be described by their transmittance and reflectance spectra (see Fig. 1(a)). From the spectra, not only we can get their negative refractive index regions, their effective permittivity and permeability[9,10], but also we can determine their resonant frequencies. We have two resonant frequencies in metamaterials, electric and magnetic resonant frequency. These frequencies are in the maximum values of the absorption curve. We show that the amplified polarization currents in the magnetic resonant frequency can be used to generate second and third harmonic signals in two different polarizations.

The highest amount of polarization currents in the metamaterials occurs in their magnetic resonant frequency. In the magnetic resonance, the incident magnetic field induces a current loop in the metamaterial (see Fig. 1 (b)). This situation can also be visualized as an LC circuit, which is operating at its resonant frequency (see the inset of Fig. 1(b)). The metallic parts from a loop which acts like an inductance and the gaps act like capacitors. In the resonant frequency of LC circuits, the impedance reaches its minimum and so the highest amplitude of currents can be seen[11], meaning that a strong current loop will be formed. Now, we show that how this current loop generates nonlinear effects in two polarizations. We divide the current loop into two parts. Upper and lower currents, and left and right currents. As mentioned above, the polarization currents act like the sources for nonlinear effects. Each of polarization currents creates a nonlinear field, perpendicular to the other one. Although the currents in these parts and the corresponding generated fields are antiparallel (e.g. Jy left and right), the added phase factor ($e^{ik_0 d}$) caused by their separation leads to a constructive field.

Nanostrips are previously introduced as the concept of metamaterials exhibiting resonant behavior in the visible frequencies range[12]. One of their distinctive properties is that we can use the initial wave's magnetic field to excite the magnetic dipoles in the magnetic resonance.

This approach induces much stronger currents in the metamaterial in comparison to the approach that uses the electric field to excite the magnetic dipole[13] (e.g. for planar SRRs). This means that, if we intend to use the magnetic field to excite the magnetic dipole in SRRs, either they should be standing SRRs, or if they are parallel to the substrate, we should use an inclined radiation, so we can have a magnetic field component normal to the SRR cross section[14]. In the latter way, only a fraction of power is injected to the metamaterial, and consequently, we have smaller nonlinear effects. However, in the case of nanostrips, we can easily excite them with normal incident radiation with a TM polarization.

To show the effect of currents, we measure the average currents on four lines on the upper ($J_{x,up}$) and lower ($J_{x,down}$) metallic parts, and left ($J_{y,left}$) and right ($J_{y,right}$) metallic parts of the nanostrips. Each of these currents generates a nonlinear field. Although each of these two currents are anti-parallel, the separation between them leads to a constructive field. To prove this claim, we measure two quantities. First of all, we measure the currents in four lines (see Fig. 1(b)). The separation between the horizontal lines is $d_1$ and the separation between the vertical lines is $d_2$. Now, we measure the following two quantities:

$$J_{x,tot} = J_{x,up} \times \exp(-ik_0 d_1) + J_{x,down} \tag{1}$$

$$J_{y,tot} = J_{y,left} \times \exp(-ik_0 d_2) + J_{y,right} \tag{2}$$

As can be seen from Fig. 2 (a) and (b), although the individual currents have maximums in different frequencies compared to the nanostrips itself, but $J_{x,tot}$ and $J_{y,tot}$ have a maximum in the same frequency as the nanostrips itself. These two perpendicular currents are responsible for creating the nonlinear effects.

Now, we discuss the required equations which are necessary for the simulating of nonlinear behavior of metals. The linear behavior of noble metals at different frequencies is described by Drude model, which is based on the equation of motion of free electrons in metals[15]:

$$m\frac{\partial^2 \vec{r}(t)}{\partial t^2} + m\Gamma \frac{\partial \vec{r}(t)}{\partial t} = -e\vec{E_0}e^{-i\omega t} \tag{3}$$

The right-hand side term is the driving force caused by the exciting wave, and the middle term is the damping force. By substituting $\vec{P(t)} = ne\vec{r}(t)$ and $\vec{J_p} = \frac{\partial \vec{P}}{\partial t}$ we have:

$$\frac{\partial \vec{J}_p}{\partial t} + \Gamma \vec{J}_p = \varepsilon_0 \omega_p^2 \vec{E} \quad (4)$$

where $\omega_p = \sqrt{\frac{n_0 e^2}{\varepsilon_0 m}}$ is the plasma frequency. The above equation is solved simultaneously with Maxwell equations followed, and then we have the linear behavior of noble metals at different frequencies:

$$\frac{\partial \vec{B}}{\partial t} = -\vec{\nabla} \times \vec{E} \quad (5)$$

$$\frac{\partial \vec{E}}{\partial t} = c^2 \vec{\nabla} \times \vec{B} - \frac{\vec{J}_p}{\varepsilon_0} \quad (6)$$

In order to consider nonlinear effects, we should also add three physical effects to the equation (3), which are the electric and magnetic components of Lorentz force and the convective derivative of the electron-velocity field[16]:

$$\frac{\partial \vec{J}_p}{\partial t} = -\Gamma \vec{J}_p + \varepsilon_0 \omega_p^2 \vec{E} + \sum_k \frac{\partial}{\partial r_k}\left(\frac{\vec{J}_p J_{pk}}{\omega_p^2 m_e \varepsilon_0 / e - \rho}\right) - \frac{e}{m_e}\left[\rho \vec{E} + \vec{J}_p \times \vec{B}\right] \quad (7)$$

In result, the final set of the equations are (5), (6) and (7). As can be seen from the above equation the nonlinear effects originate from the polarization currents.

**Model description**

The proposed design consists of five layers. 10nm thick alumina ($Al_2O_3$) on the top and the bottom of the structure, and a gold-alumina-gold in the middle of the structure (see Fig. 3 (a)). The thickness of gold layers is 35nm and the thickness of alumina layer is 40nm. The proposed structure of the nanostrips lies upon a glass substrate. For the simulations, we use COMSOL Multiphysics, which is a finite element based simulation software. As for the simulation parameters, continuity boundary conditions on the left and right boundaries and non-reflecting conditions on the top and bottom boundaries are used. We excite the model with a Gaussian wave with TM polarization from the top boundary. The optical parameters in the simulation are n=1.5 for glass, n=2.25 for alumina[17], $\omega_p$ = 13.67×10$^{15}$ 1/s, $\Gamma$=0.0648×10$^{15}$ 1/s, $\varepsilon_\infty$=9.84 for gold[15,18].

For measuring SH and TH signals, two probes are used at the beginning and at the end of the simulation frequency, measuring the reflected and transmitted wave. The reflected wave shows the same behavior with lower amplitude. The resulting harmonic signals are shown in Fig. 3(b). These curves are obtained by measuring the amplitude of Fourier transformation of the transmitted wave. As can be seen form Fig. 1 (b), the nanostrips have inversion symmetry in x-direction, and no inversion symmetry in y-direction. Thus, $J_{x,tot}$ generates a third harmonic signal and $J_{y,tot}$ generates a second harmonic signal. In other words, we have a third harmonic signal in the x direction, same as the initial field, and a second harmonic signal in the y direction. In SRRs structures[19], a similar experimental result which can be explained by our analysis for nonlinear signal generation in the magnetic metamaterials was reported and it supports our claim.

**Discussion**

In this paper, we investigated the nonlinear behavior of magnetic metamaterials as an ultrathin nonlinear optical component. We have shown how magnetic metamaterials generate harmonic signals in two polarizations. We discussed how the current loop induced in the magnetic resonant frequency of the metamaterial can be divided into two orthogonal parts, and how these parts generate harmonic signals in two polarizations. We also showed how the symmetric properties of the metamaterial determine the type of generated harmonic signal.

**Acknowledgements:**

This work was supported by Young Investigator Research program (No. 2015R1C1A1A02036464) and Engineering Research Center program (No. 2015R1A5A1037668) through the National Research Foundation of Korea (NRF) grant funded by the Ministry of Science, ICT and Future Planning (MSIP) of Korean government.

**Author contributions:**

I.S. and J.R. initiated and developed the idea. I.S. performed the theoretical calculations. J.R and I.K. provided the information for design and fabrication. I.S. and J.R. prepared the manuscript. A.Z. and J.R. guided the research. All authors contributed to discussions.

**Competing interests**

The authors declare that they have no competing interests.

**Figures**

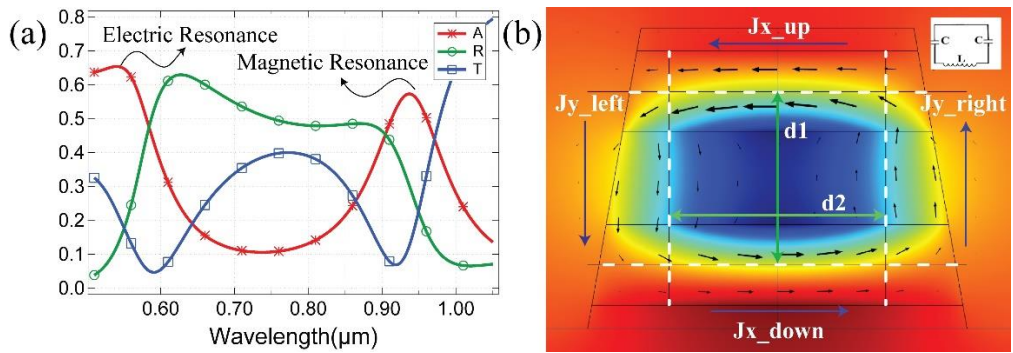

Fig. 1. (a) Transmission, reflection and absorption spectra of nanostrip. The peaks in the absorption curve determine the resonant frequencies. (b) Magnetic field distribution and electric field displacement (black arrows) in magnetic resonant frequency. The external magnetic field induces a current loop which can be divided into two orthogonal groups. The current loop components are measured on the white dashed lines. The separation between these lines are shown by $d_1$ and $d_2$. The inset shows the equivalent circuit.

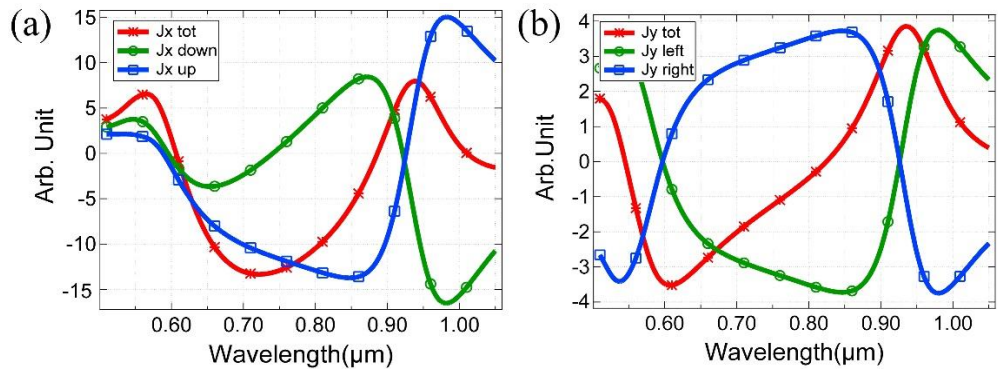

Fig. 2. (a) Polarization currents in the upper and lower metallic parts ($J_x$ up and down), and the $J_{x,tot}$. (b) Polarization currents in the left and right metallic parts ($J_y$ left and right), and the $J_{y,tot}$. Both components of $J_{tot}$ has the same maximum as the absorption curve in Fig. 1 (a).

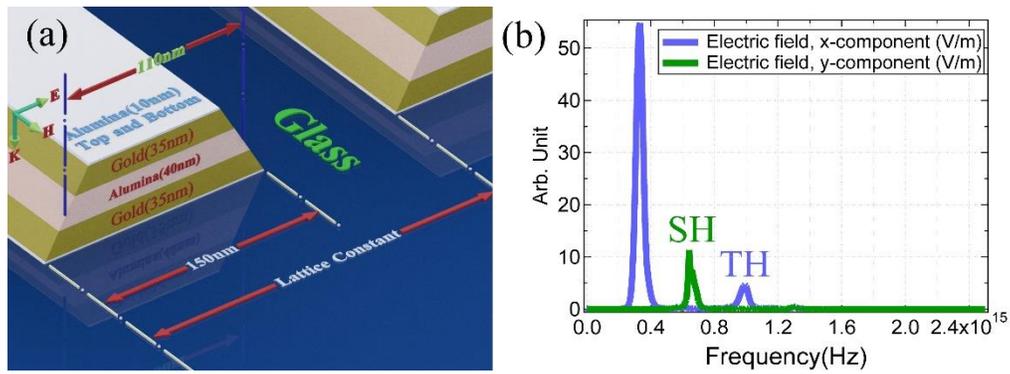

Fig. 3. (a) Geometrical parameters of the model. The model is a 5-layer trapezoid, which is assumed to be very long. (b) The fourier transform of the transmitted signal. The initial wave and the third harmonic signals are in the x-direction and the second harmonic signal is in the y-direction.